# Design of cycloidal rays in optical waveguides in analogy to the fastest descent problem

XIONG HUI*[1], ZIJUN HE[1] AND YANGJIE LIU (刘泱杰) [2,1,3]

[1]*School of Microelectronics, Hubei University, Wuhan 430062, China*

[2]*School of Physics, Hubei University, Wuhan 430062, China*

[3]*Lanzhou Center for Theoretical Physics, Key Laboratory of Theoretical Physics of Gansu Province, Lanzhou University, Lanzhou 730000, China*

*Corresponding author: xionghui@fudan.edu.cn

**In this work, we present the design of cycloidal waveguides from a gradient refractive index (GRIN) medium in analogy to the fastest descent problem in classical mechanics. Light rays propagate along cycloids in this medium, of which the refractive index can be determined from relating to the descending speed under gravity force. It can be used as GRIN lenses or waveguides, and the frequency specific focusing and imaging properties have been discussed. The results suggest that the waveguide can be viewed as an optical filter. Its frequency response characteristics change with the refractive index profile and the device geometries.**

## 1. Introduction

The connection between classical mechanics and optics is a long lasting and still intriguing problem. The movement of an object under the action of an arbitrary force, as dictated by the Euler-Lagrange equation, appears to be deeply connected with the propagation of light in a special class of medium with gradient refractive index (GRIN), as dictated by the famous Fermat's principle. Both follow the least action principle, which can be described by the variational method. Guided by this beautiful analogy, people formulated the peer law of Newtonian mechanics in optics [1], developed general methods of ray tracing in a symmetric GRIN profile from Fermat's ray invariants [2,3] and optical Binet equation [4], and solved the optical Lagrange equation using the first integral method to control light trace in GRIN media [5].

In recent years, progress in material science has made it possible to fabricate optical medium of GRIN type by a variety of techniques, such as ion-exchange process [6,7], photo-thermal reaction in glass [8], nano-scale porous silicon [9], gradient-structured metamaterials [10], and additive manufacturing with nanocomposite materials [11,12]. These artificial materials enable people to apply the well-known methods and conclusions in mechanics in novel optical design, which has potential application in invisibility cloaks [13,14], perfect imaging lenses [15,16], and transformation optical media [17,18].

In this work, we propose the design of a GRIN optical device based on the analogy to the famous problem of fastest descent (also known as the Brachistochrone problem). The light paths follow cycloidal curves, just like the descending object, if the refractive index profile is devised to imitate the effect of gravitational force acting on the object. In this medium, we can construct the refractive index profile to be cylindrically symmetric, such that light is confined transversally and directed along the cylinder axis, which can be applied as a waveguide. Furthermore, we provide field simulation results to support our design. The light field distribution reveals the physical nature behind the geometric optics that light rays obey in the medium.

## 2. Theoretical model

To restate the fastest descending problem, we need to seek a track for an object to descend between two fixed points within the shortest time. In Cartesian coordinates, the object descends at a speed under the action of gravitation $v = \sqrt{2g_0 y}$, where $g_0$ is the gravitational acceleration. Therefore, the duration time for the object to descend from point A to point B is:

$$\tau = \int_A^B \frac{\sqrt{1+f'(y)^2}}{\sqrt{2g_0 y}} dy, \tag{1}$$

where the actual path is determined by the function $x = f(y)$. Through variational method, the path that minimizes $\tau$ is a cycloid, which satisfies the parametric equations:

$$x = \frac{b}{2}(\theta - \cos\theta), y = -\frac{b}{2}(1 - \cos\theta) \tag{2}$$

in which $b$ is a parameter determining the spatial extension of the curve. The schematics of a cycloid is shown in Fig. 1(a). When it comes to optics, the speed of light propagating in a dielectric medium is $v = c/n$, where $n$ is the refractive index of the medium. It is found that if we prescribe a refractive index profile of

the form:

$$n(y) \propto \frac{1}{\sqrt{-c_0 y}} \quad (y < 0), \qquad (3)$$

where $c_0$ is a constant, the medium plays a similar role as gravitation does in the fastest descending problem. Consequently, light rays will simply follow cycloids described by Eq. (2), which minimize the length of optical path for the light rays. This fact agrees with Fermat's principle. On the other hand, light rays can only propagate along cycloids when light feels a GRIN profile of the medium, as dictated by Eq. (3). It represents a specific class of GRIN media, and we call them the cycloidal media in following discussion for convenience.

## 3. Results and discussion

### 3.1 Constant gravitation model

In Fig. 1(b), we depict the trajectories of light rays in the cycloidal medium with $c_0 = 1 g_0$/mm. Light emits at different angles, and each path is a cycloid from the same starting point at $l = 0$, where $l$ denotes the distance light goes in the $x$-direction. If we reverse the light propagation, the paths end with convergence, showing that parallel light rays are focused to the starting point, i.e. the image of an infinitely distant object. Therefore the medium itself can be used as a lens, with its focus length depending on the incident position of light rays.

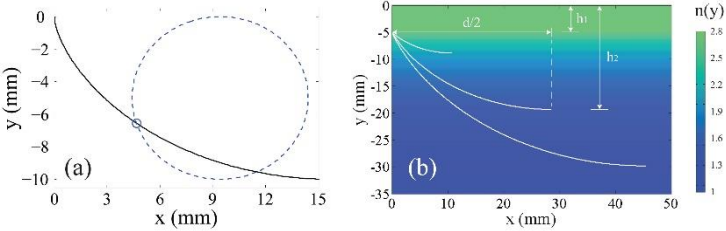

Fig. 1 (a) Cycloidal track in the fastest descending problem, which is the trace of a point on a rotating circle. (b) Light trajectories in white curves in the cycloidal medium. The notations of $h_1$, $h_2$ and $d$ mentioned in the main text are labeled for one of the curves. On the background mesh colors are used to indicate the value of refractive index.

If we rotate the cycloidal medium around the $y = 0$ axis, we can construct a cylindrical waveguide. Fig. 2 shows its cross-section, in which light propagation is restricted in-plane due to symmetry in the azimuthal direction. We can also insert a uniform layer with constant refractive index in the center, serving as the core layer of the waveguide.

Imagine that light enters the waveguide from left. Without losing generality, we assume light enters in the lower half side of the waveguide at $y = -h_1$. Starting from the entrance point, the light rays follow cycloidal trajectories, propagating down from inner to peripheral layer of the waveguide until it reaches $y = -h_2$, at which the light rays become parallel to the horizontal axis. Then it is deflected back to the inner layer, and goes straight through the core layer where $n(y)$ is a constant.

Due to the rotation symmetry of the refractive index profile, the light path in the upper half side for $y > h_1$ is symmetric to that for $y < -h_1$. In this way, the propagation of light in the waveguide is periodic, where the period length $d$ depends on the entrance angle of each light ray, denoted as $\alpha$. We plot the first few periods of light trajectories with different $\alpha$ in Fig. 2. If absorption loss is neglected, the light just continues to propagate in the waveguide and can be guided to the far end of the device.

Among all the light rays, there are a group of rays with periods that satisfy rational ratios between each other. In this case, we can find a least common multiple of periods for these rays, so they constitute a commensurate group. After propagating for multiple periods, these light rays will arrive at the same point, similar to the focusing behavior of an ordinary lens.

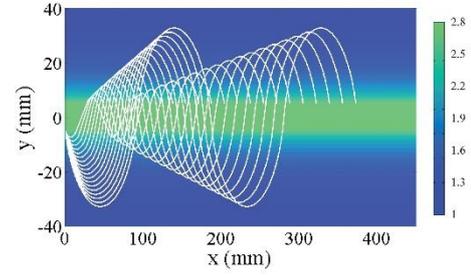

Fig. 2 Cycloidal trajectories of light rays (white curves) within the first two periods in the waveguide under the constant gravitation model. The entrance angle $\alpha$ ranges from -66.7° to -33.9°, and the distance $d$ that light travels in one period differs with $\alpha$.

However, unlike the fact that there are infinitely many light rays that are focused for the ordinary lens, the number of light rays in the cycloidal waveguide that arrive at the same focus point is limited. For example, let us examine all the light rays with the entrance angle $\alpha$ from about -70° to -20°. The variation of $d$ versus $\alpha$ in this range is plotted in Fig. 3(a). As the data shows, the ratio between the maximum and minimum of $d$ for these rays is about 12, and therefore the possible rational ratios in this condition are:

$$r_m = \frac{m \cdot d_{min}}{d_{max}} = \frac{m}{12}, (1 \leq m \leq 12) \qquad (4)$$

which includes periods:

$$d = (\tfrac{1}{12}, \tfrac{1}{6}, \tfrac{1}{4}, \tfrac{1}{3}, \tfrac{5}{12}, \tfrac{1}{2}, \tfrac{7}{12}, \tfrac{2}{3}, \tfrac{3}{4}, \tfrac{5}{6}, \tfrac{11}{12}, 1) \times d_{max} \qquad (5)$$

It contains 12 elements in total, each corresponding to a certain $\alpha$, where $d_{max}$ and $d_{min}$ are the maximum and minimum of period in the group with the largest and smallest $|\alpha|$ respectively. For the rays we select, $d_{max} = 97.4$ mm, $\alpha = -66.7°$ and $d_{min} = 8.12$ mm, $\alpha = -21.2°$. Given those ratios in Eq. (5), the least common multiple of the numerators is 2310, which means that at $l = 2310 d_{max}$ all these light rays are converged together after propagating for a multiple number of periods. In addition, prior to this point there are other positions at which only part of the light rays are converged. For instance, at $l = 30d_{max}$, the light rays with the periods:

$$d = (\frac{1}{12}, \frac{1}{6}, \frac{1}{4}, \frac{1}{3}, \frac{5}{12}, \frac{1}{2}, \frac{2}{3}, \frac{3}{4}, \frac{5}{6}, 1) \times d_{max} \quad (6)$$

are converged, and at $l = 6d_{max}$, the combination becomes:

$$d = (\frac{1}{12}, \frac{1}{6}, \frac{1}{4}, \frac{1}{3}, \frac{1}{2}, \frac{2}{3}, \frac{3}{4}, 1) \times d_{max} \quad (7)$$

It is evident that the position of the focus point depends on the selection of a certain combination of rational ratios. If we place an ordinary lens behind the waveguide, it will form the image of those light rays from a selected focus point. From the viewpoint of Fourier optics, each light ray represents a certain spatial frequency component. The refraction and imaging of light rays through a lens corresponds to a superposition of different spatial frequency components. It can be described by the inverse discrete Fourier transform with finite items:

$$I(x,y) = \frac{1}{(2\pi)^2} \sum_{i,j} F(i,j) e^{i(\omega_i x + \omega_j y)} \Delta \omega_i \Delta \omega_j, \quad (8)$$

in which $I(x,y)$ is the light intensity distribution on the image plane, and $F(i,j)$ is the discrete frequency spectrum of the transmission function on the object plane. From the Kirchhoff diffraction formula of a thin lens [19], the spatial frequency of an object under coherent illumination equals to the tangent of the entrance angle $\alpha$ multiplied by the wave number $k_0$ of light:

$$\omega_i = k_0 \tan \alpha_i, \omega_j = k_0 \tan \alpha_j \quad (9)$$

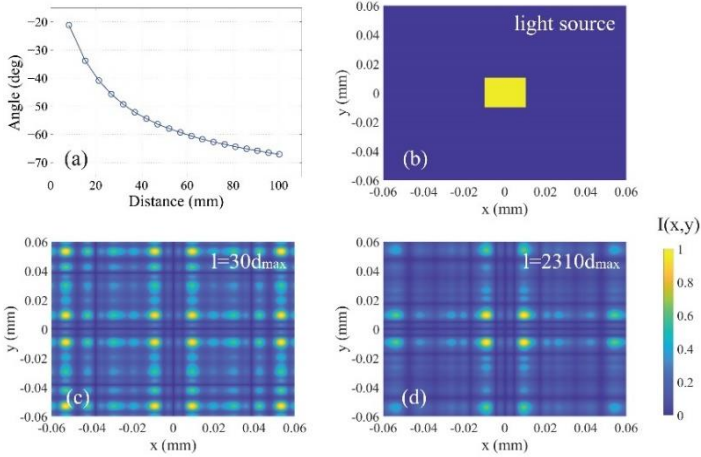

Fig. 3 (a) Entrance angle $\alpha$ versus period length $d$ in constant gravitation model. (b) Geometry of the light source at $l = 0$. (c) and (d) Normalized images reconstructed from finite sum of spatial frequency components in the commensurate group at $l = 30d_{max}$ and $2310d_{max}$, respectively.

where the index $i,j$ run over the items in the commensurate group, and $\Delta\omega_{i,j}$ in Eq. (8) are intervals of discrete frequencies related to sampling precision, which can be treated as constant coefficients. In this sense, the waveguide acts as a spatial frequency selector along its axis, as certain positions on the axis correspond to certain combinations of spatial frequency components.

On the Fourier plane of the lens, these components overlap to form a filtered image of the light source, which may be distinct from the original appearance. To shed light on the filtering effect, we assume that light is emitted from a coherent light source of square shape, as shown in Fig. 3(b). It is worth noticing that, in order to observe the frequency specific image in experiments, an iris has to be used and placed at the focus point, blocking out all other rays that do not belong to the commensurate group.

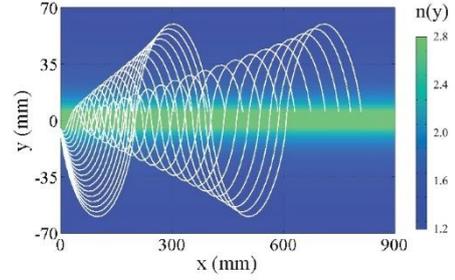

Fig. 4 Cycloidal trajectories of light rays (white curves) within the first two periods in the waveguide under the variable gravitation model. The entrance angle $\alpha$ ranges from -68.8° to -29.2°.

The results at $l = 30d_{max}$ and $2310d_{max}$ are shown in Fig. 3(c) and 3(d), which tell us about the evolution of imaging properties in frequency domain. When $l$ is increased, there are more spatial frequency components from the light source making a contribution, and the image in Fig. 3(d) resembles the source profile to a greater extent than that in Fig. 3(c), as reflected in the ratio $I_{max}^c/I_{max}^e$ taken between local intensity maximum in the central area and near the edge on the image plane. The result is 0.989 in Fig. 3(c) and 1.542 in Fig. 3(d), which shows that light intensity is increasingly distributed around the central area as $l$ increases.

Since the selected light rays emit at relatively large angles, they represent high frequency components of spatial structure. In Fourier transform, spectrum in high frequency range comes from rapid change in space domain. Consequently, it is the edge of the light source that is highlighted in the image, while the clarity varies at different distances, as the number of spatial frequency components included in the commensurate group changes. From another perspective, the image can be understood as the interference pattern between light rays arriving at the image plane with different wavefronts.

### 3.2 Functional gravitation model

In the problem of fastest descent, gravitational force is treated as a constant. If the gravity field changes, the extremal track of the descending object does change its path. This reminds us that for optical design, we can extend the model to introduce space dependent gravitation generated by either real or artificial mass distribution, under which $g(y)$ is an arbitrary function of $y$. As a more general result, the trajectories of light rays follow the equation:

$$x = \text{sign}(h_2 - h_1) \int_{h_1}^{h_2} \sqrt{-\frac{\int_0^y g(\gamma)d\gamma}{\int_0^y g(\gamma)d\gamma + b}} dy, \quad (10)$$

For instance, assuming $g(y)$ is a quadratic function of $y$:
$$g(y) = c_1 y^2 + c_2 y + c_3, \quad (11)$$

In this case, $n(y)$ takes the form:
$$n(y) \propto \frac{1}{\sqrt{-\frac{1}{3}c_1 y^3 - \frac{1}{2}c_2 y^2 - c_3 y}}, \quad (12)$$

The denominator is the square root of a polynomial of variable $y$ with parameters $c_1$, $c_2$ and $c_3$. We can change $c_1$, $c_2$ and $c_3$ to modify the refractive index profile, and hence changing the light trajectories. It is found that for certain values of $c_1$, $c_2$ and $c_3$, the difference of period $d$ between light rays emitting at different $\alpha$ becomes larger, such that the range of $d_{max}/d_{min}$ becomes larger. For example, when $c_1 = 1.2 \times 10^{-4} g_0/mm^3$, $c_2 = 1.7 \times 10^{-2} g_0/mm^2$ and $c_3 = 0.82 g_0/mm$, $d_{max}/d_{min}$ can reach 24. As Fig. 4 shows, light follows trajectories that are quasi-cycloidal, and in this case the commensurate group becomes:

$$r_m = \frac{m \cdot d_{min}}{d_{max}} = \frac{m}{24}, (1 \leq m \leq 24) \quad (13)$$

Compared to the case when $d_{max}/d_{min} = 12$, the number of light rays with rationally harmonic frequencies increases. The focus point shifts approximately to $l = 2.2 \times 10^8 d_{max}$ for all these light rays to meet together. In fact, we need not to go as far as the full focus length, since most rays already encounter each other at a smaller distance. The corresponding imaging results at $l = 6d_{max}$, $30d_{max}$ and $2310 d_{max}$ are shown in Fig. 5.

$l = 6d_{max}$ the imaging result already resembles the patterns that obtained at a much longer distance of $l = 30 d_{max}$ in the constant gravitation model, as evidenced by comparison between Fig. 3(c) and Fig. 5(b), in which $I_{max}^c/I_{max}^e = 0.989$ and $1.067$ respectively. The result illustrates that the waveguiding properties of the cycloidal medium can be manipulated through changing the functional form of $g(y)$, in analogy to changing the gravitational force in the fastest descent problem.

The partial imaging properties of the cycloidal waveguide is reminiscent of an important self-imaging phenomenon in wave optics, the fractional Talbot effect in Fresnel diffraction [20]. The Talbot image is also space dependent, consisting of a series of superposed images of a periodic object. Each image represents a different subset of spatial frequency components of the object.

### 3.3 Field simulation result

Fig. 6 shows the field simulation results under the constant and the functional gravitation model at the wavelength of 500μm and 200μm respectively. It reveals how light propagates in the cycloidal medium from wave optics perspective. A plane wave of width 1.5mm and initial field magnitude of 1V/m is incident from the left side. The wavefront then follows the eikonal equation and gets distorted by the GRIN profile on the light path. Finite element method was used to calculate the static distribution of transverse electric field when light passes through the cycloidal medium. Perfectly matching layers were assigned to the boundaries. Note that compared with Fig. 2 and Fig. 4, the figure has been rotated by 45° for the convenience of setting up boundary conditions.

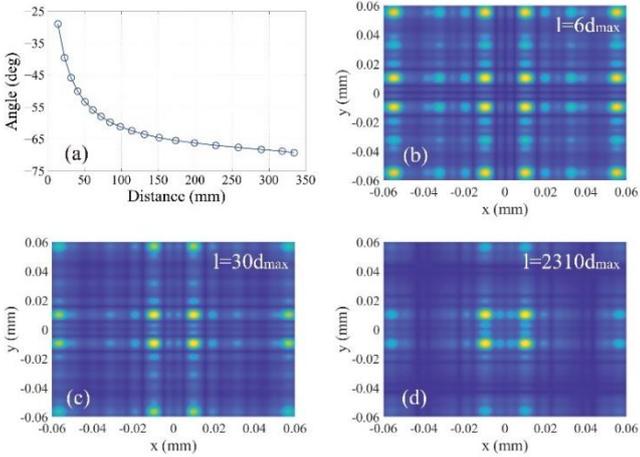

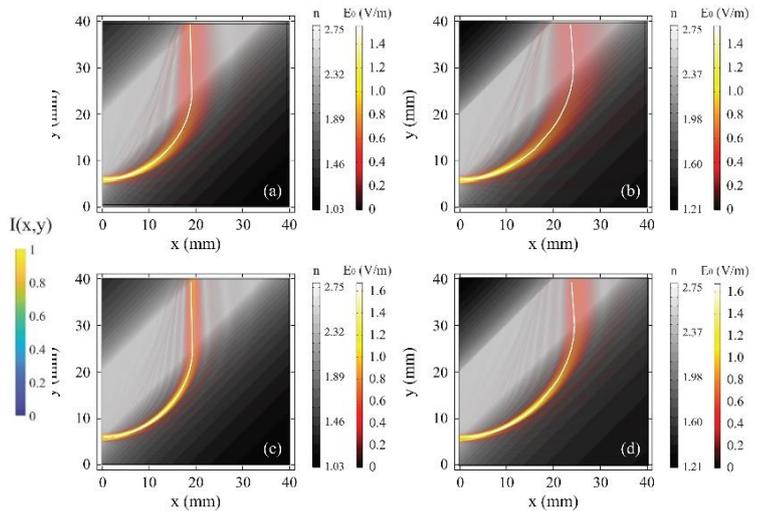

Fig. 5 (a) Entrance angle $\alpha$ versus period length $d$ in functional gravitation model. (b)-(d) Normalized images reconstructed from summing up finite items of spatial frequency components in the commensurate group at $l = 6d_{max}$, $30d_{max}$ and $2310d_{max}$, respectively.

It is seen that the edge of the light source becomes sharper compared with Fig. 3 for the same $l$, as more spatial frequency components convey more information in frequency domain to the imaging result in the functional gravitation model. Consequently, at

Fig. 6 Electric field of plane waves propagating in the cycloidal medium under the constant gravitation model (a, c) and the variable gravitation model (b, d). The wavelength of input light is 500μm (a, b) and 200μm (c, d) respectively. To distinguish electric field from the background substance, we use thermal color scale to represent the electric field magnitude, and gray scale to represent the refractive index.

It is found in Fig. 6 that the maximum of the electric field magnitude follows a deflected trace as the refractive index changes gradually, which coincides largely with the corresponding cycloids. Moreover, it is observed that as the wavelength decreases to 200μm, the propagation of light in the waveguide shows less scattering and dispersion loss, forming a farther-reaching trace along the light path obtained in rays tracing. This feature is in accordance with the fact that geometrical optics becomes a better approximation when the wavelength decreases, wherein the eigen-modes of the waveguide reduce to light rays, as described by cycloidal curves.

In general, the concept of geometrical optics is well established for GRIN media if the wavelength of light satisfies $|\nabla \lambda| \ll 1$, which holds true in our model. It asserts that the spatial variation of wavelength should be small enough. The results imply that Fermat's principle together with its analogy to classical mechanics is an instructive clue to modeling and designing the proper refractive index profile given the path of light rays. It provides us with an ancient but powerful tool, bringing to optical design more flexibilities and possibilities.

With respects to fabrication techniques of the waveguide proposed in this work, the broad index range the GRIN profile covers is challenging for the fabrication process. Future research is called for achieving a large refractive index span of the GRIN profile in optical materials, which allows for expanding the design space of GRIN optical elements and devices.

## 4. Conclusion

In this work, we transplant the classical problem of fastest descent into optics, which enables the design of a GRIN medium in which light rays propagate along cycloids. The refractive index of this medium can be prescribed in analogy to the descending speed under gravitation. With rotational symmetry, the medium can function as a waveguide, and its light transmitting, focusing and imaging properties have been discussed.

If we further change the functional form of the refractive index, the trajectories of light rays can be controlled, hence altering the ratio between the distance light rays propagate in one period. In this way, the number of rational ratios in the commensurate group changes, leading to different combinations of light rays at different distances, so the waveguide functions as an optical filter. Specific spatial frequency components of the light source can be guided and selected out for further utilization. Field simulation results support the geometrical optical properties obtained in our method.

## Acknowledgements

The authors would like to thank Hubei University for funding support. This work was also supported by the National Science Foundation of China for Young Scholars (no. 11804087), the National Natural Science Foundation of China (no. 12047501), the Educational Commission of Hubei Province of China (no. T2020001) and the Science and Technology Department of Hubei Province of China (no. 2018CFB148) granted to Y. Liu.

## Disclosure

The authors declare no conflicts of interest.

## Data availability

Data underlying the results presented in this paper are not publicly available at this time but may be obtained from the corresponding author upon reasonable request.